\documentclass[runningheads]{llncs}
\usepackage{graphicx}
\usepackage{xcolor}
\usepackage{multirow}

\newcommand\blfootnote[1]{%
  \begingroup
  \renewcommand\thefootnote{}\footnote{#1}%
  \addtocounter{footnote}{-1}%
  \endgroup
}

\begin{document}
%
\title{Bringing Cognitive Augmentation to Web Browsing Accessibility}
%
%
%
\author{Alessandro Pina\inst{1} \and
Marcos Baez\inst{2} \and
Florian Daniel\inst{1}}

\authorrunning{A. Pina et al.}
%
\institute{Politecnico di Milano, Milan, Italy \\
\email{alessandro.pina@mail.polimi.it}, \email{florian.daniel@polimi.it} \and
LIRIS – University of Claude Bernard Lyon 1, Villeurbanne, France \\
\email{marcos.baez@liris.cnrs.fr}}
\maketitle              
\begin{abstract}
In this paper we explore the opportunities brought by cognitive augmentation to provide a more natural and accessible web browsing experience. 
We explore these opportunities through \textit{conversational web browsing},  an emerging interaction paradigm for the Web that enables blind and visually impaired users (BVIP), as well as regular users, to access the contents and features of websites through conversational agents.
%
Informed by the literature, our previous work and prototyping exercises, we derive a conceptual framework for supporting BVIP conversational web browsing needs, to then focus on the challenges of automatically providing this support, describing our early work and prototype that leverage heuristics that consider structural and content features only.

\end{abstract}
\section{Introduction}
Accessing the Web has long relied on users to correctly process and interpret visual cues in order to have a proper user experience. Web browsers as well as information and services on the Web are optimised to make full use of user's visual perceptive capabilities for organising, delivering and fulfilling their goals. 
This introduces problems for blind and visually impaired people (BVIP) who due to genetic, health or age-related conditions are not able to effectively rely on their visual perception \cite{barreto2019visual}.\blfootnote{This is a post-peer-review, pre-copyedit version of an article accepted to the International Workshop on AI-enabled Process Automation, at ICSOC 2020.}

\begin{figure}[h]
\centering
  \includegraphics[width=.9\columnwidth]{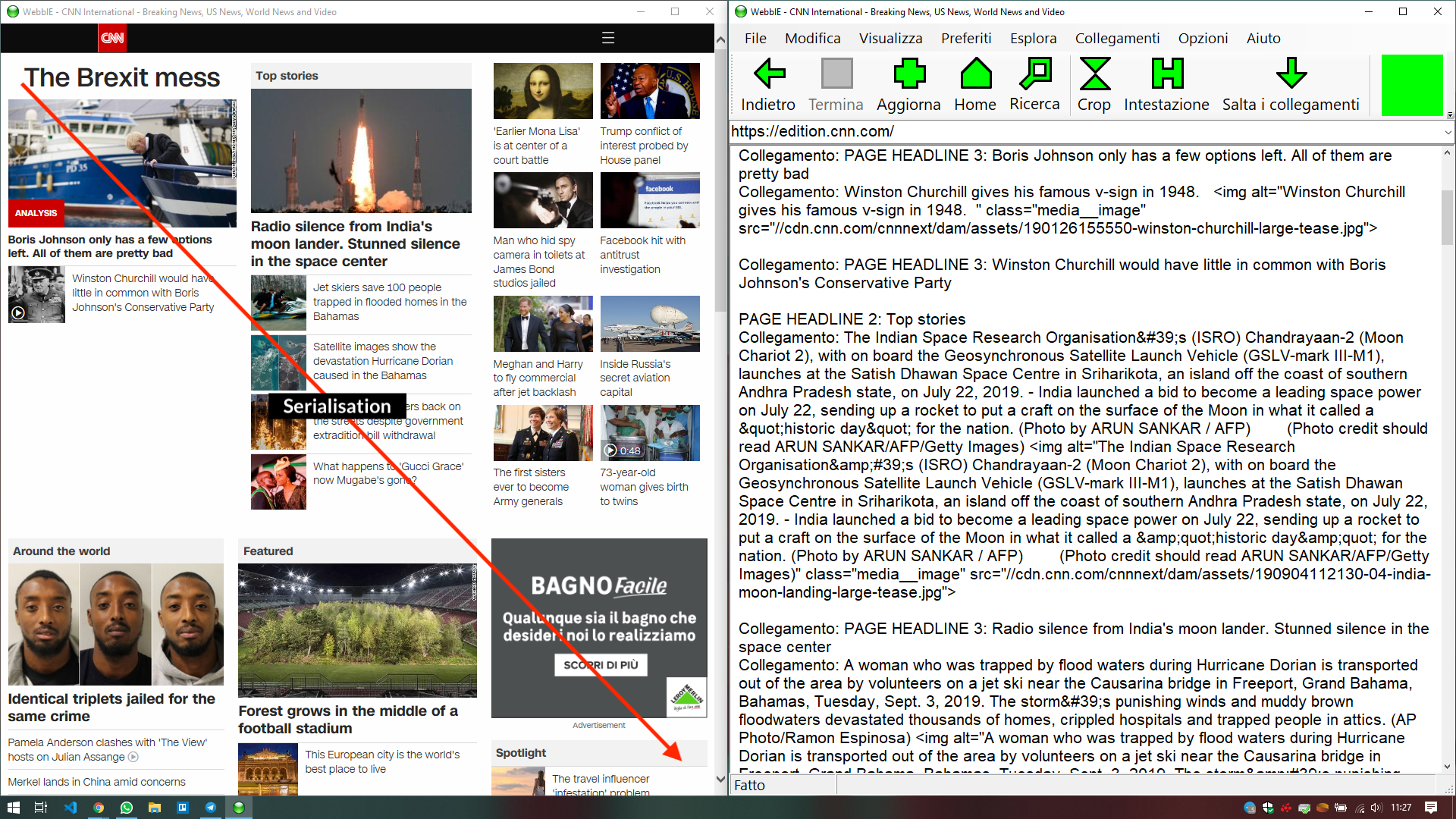}
  \caption{Example serialisation of a website with by a screen reader. HTML elements are read typically from top to bottom, as informed by the website HTML structure }
  \label{fig:screen-reader}
\end{figure}

Assistive technology such as screen readers have traditionally supported BVIP users in interacting with visual interfaces. These tools exploit the \textit{accessibility tags} used by Web developers and content creators in order to facilitate access to information and services online, typically by reading out the elements of the website sequentially from top to bottom (see Figure \ref{fig:screen-reader}). They are usually controlled with a keyboard, offering shortcuts to navigate and access content. The challenges faced by BVIP in browsing the Web with this type of support is well documented in the literature, ranging from websites not designed for accessibility \cite{lazar2007frustrates,bigham2017effects} to limitations of screen reading technology \cite{mahmud2007csurf,zhu2010sasayaki,ashok2015capti}.

Cognitive augmentation has been regarded as a promising direction to empower populations challenged by the traditional interaction paradigm for accessing information and services \cite{barukh2020}. Conversational browsing is an emerging interaction paradigm for the Web that builds on this promise to enable BVIP, as well as regular users, to access the contents and services provided by websites through dialog-based interactions with a conversational agent \cite{baez2019conversational}. Instead of relying on the sequential navigation and keyboard shortcuts provided by screen readers, this approach would enable BVIP to express their goals by directly ``talking to websites". The first step towards this vision was to identify the conceptual vocabulary for augmenting websites with conversational capabilities and explore techniques for generating chatbots out of websites equipped with bot-specific annotations \cite{chitto2020automatic}.  


In this paper we take a deeper dive into the opportunities of cognitive augmentation for BVIP by building a conceptual framework that takes the lessons learned from the literature, our prior work and prototyping exercises to highlight areas for conversational support.
We then focus on the specific tasks that are currently served rather poorly by screen readers, and describe our early work towards a \textit{heuristic-based} approach that would leverage visual and structural properties of websites to translate the experience of graphical user interface into a conversational medium.


\section{Conceptual framework}
\subsection{Motivating scenario}
Before introducing the main concepts, we illustrate our vision by describing an example interaction of a BVIP looking for information on COVID-19 on a local newspaper. 
Peter, 72, is a visually impaired man affected by Parkinson disease,  who keeps hearing about the new
virus COVID-19 on TV and wants to be updated constantly about the recent news from his favorite local newspaper, The Tambury Gazette. However, his experience with screen readers has been poor and frustrating, often requiring assistance from others to get informed.

\begin{figure}[t!]
\centering
  \includegraphics[width=\columnwidth]{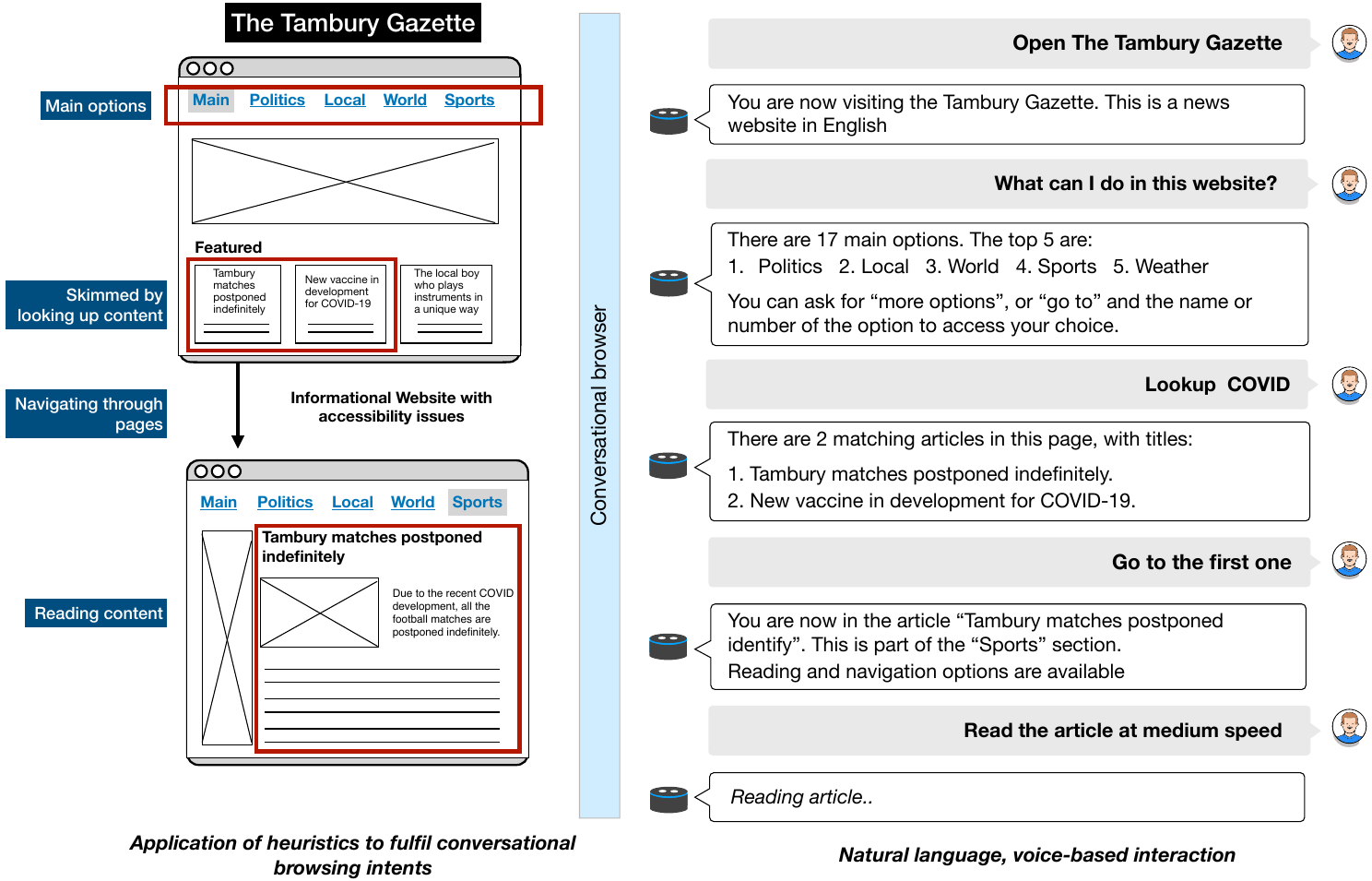}
  \caption{Example conversational browsing on an information-intensive website. }
  \label{fig:concept}
\end{figure}

The vision is to enable users like Peter to browse the Web by directly ``talking" to websites. As seen in Figure \ref{fig:concept}, the user interacts with the website in dialog-based voice-based interactions with a conversational agent (e.g., Google Assistant). 
The user can start the session by searching for the website, or opening it up directly if already bookmarked. Once open, the user can inquire about the relevant actions that are available in its current context (e.g., \textit{``What can I do in this website”}), which are automatically derived by the conversational agent based on heuristics. Instead of sequentially going through the website, the user can lookup for specific information within the website matching his interests (e.g., \textit{``Lookup COVID"}). The user can then follow up on the list of resulting articles and chose one to be read out.
As part of these interactions, the user can use the voice commands to navigate and get oriented in the website.

The above illustrates the experience of browsing a website by leveraging natural language commands that improve over the keyboard-based sequential navigation of websites. As we will see, more advanced support can be provided by leveraging the contents and application-specific domain knowledge, but in this work we focus on improving on the features provided by screen readers, making no assumptions about compliance with accessibility and bot-specific annotations.




\subsection{Characterising conversational browsing support}

Enabling conversational browsing requires first and foremost to understand the type of support that is needed to meet BVIP needs. Informed by previous research, our own work and prototyping experiences, we highlight a few relevant areas in Table  \ref{tab:intents} and describe them below.

 \begin{table*}[ht!]
\vspace{-8pt}
\caption{Categories of support for engaging BVIP in conversations with websites}~\label{tab:intents}  
\centering
\begin{tabular}{|p{3cm}|p{2.5cm}|p{6cm}|}
\hline
\textbf{Category} & \textbf{Skills}  & \textbf{Examples} \\ \hline
\multirow{5}{3.5cm}{Metadata \& content} &  Overview & ``What is this website about?" \\ \cline{2-3}
&  Content Q\&A & ``When are sports coming back?" \\ \cline{2-3}
&  Summary & ``Summarise the article?" \\ \cline{2-3}
 &  About & ``Who are the authors of this article?” \\  \cline{2-3}
&  Yes/No & ``Is the article written in English?” \\  \hline 
\multirow{5}{3.5cm}{Browsing} &  Outline & ``What can I do in this website?” \\ \cline{2-3}
&  Orientation & ``Where am I?” \\ \cline{2-3}
 &  Navigation & ``Go to the main page"; ``Next article"
 \\ \cline{2-3}
 &  Lookup & ``Lookup COVID"
 \\ \cline{2-3}
 &  Reading & ``Read article"; ``Stop reading"
 \\ \hline 
\multirow{2}{3.5cm}{Workflows} &  Element-specific & ``Fill out the form" \\ \cline{2-3}  
 &  App-specific & ``Post a new comment on the news article" \\ \hline 
\multirow{5}{3.5cm}{Operations} &  Open & ``Open The Tambury Gazette" \\ \cline{2-3}  
 &  Search & ``Search for The Tambury Gazette"\\ \cline{2-3}  
 &  Bookmark & ``Bookmark page The Tambury Gazette" \\  \cline{2-3}  
 &  Speech & ``Increase speech rate"\\ \cline{2-3}  
 &  Verbosity & ``Turn on short interactions"\\ \hline 

\end{tabular}
\vspace{-10pt}
\end{table*}

\noindent\textbf{Conversational access to content and metadata.}
BVIP should be able to satisfy their information needs without having to sequentially go through all the website content and structure -- a process that can be time consuming and frustrating for screen reader users \cite{lazar2007frustrates}.
This support is rooted in the ongoing efforts in conversational Q\&A \cite{braun2017evaluating} and document-centered digital assistant \cite{ter2020conversations}.
The idea is to support BVIP users to perform natural language queries (NLQ) on the contents of websites, and to inquire about the properties defined in the website's metadata. For example, a BVIP user might request an \textit{overview} of the website (e.g., ``What is this website about?"), engage in \textit{question \& answering}, with questions that can be answered by referring directly to the contents of the website (e.g., ``When are the sports coming back?"), and ask for \textit{summaries} of the contents of the website, its parts or responses from the agent (e.g., ``Summarise the article"). Users might also ask \textit{about the properties} and metadata of the artefacts, such as last modification, language, authors (e.g., ``Who are the authors of this article?"), or simply engage in \textit{yes/no} questions on metadata and content (e.g., ``Is the document written in English?").

\smallskip
\noindent\textbf{Conversational browsing.} 
BVIP should be allowed to explore and navigate the artefacts using natural language, so as to support more traditional information seeking tasks. The idea is to improve on the navigation provided by traditional screen readers, which often require learning complex shortcuts and lower level knowledge about the structure of artefact (e.g., to move between different sections), by allowing users to utter simpler high level commands in natural language.  This category of support is inspired by the work in Web accessibility, in using spoken commands to interact with non visual web browsers \cite{vesnicer2003voice,ashok2015capti} 
and conversational search \cite{trippas2015spoken,trippas2017people}.
For example, BVIP should be able to inquire about the website organization and get an \textit{outline} (e.g., ``What can I do in this website?"), and \textit{navigate} through the structure of the website and even across linked webpages 
(e.g., ``Go to the main page"), and being able to get \textit{oriented} during this exploratory process (e.g., ``Where am I?"). The user should also be able to \textit{lookup} for relevant content to avoid sequentially navigating the page structure (e.g., ``Lookup COVID").

\smallskip
\noindent\textbf{Conversational user workflows.}
BVIP should also be able to enact user workflows by leveraging the features provided by the website. This is typically done by the users, enacting their plan by following links, filling out forms and pressing buttons. This low level interactions have been explored by speech-enabled screen readers such as Capti-Speak \cite{ashok2015capti}, enabling user to utter commands (e.g., ``press the cart button", ``move to the search box"). We call these \textit{element-specific} intents.
In our previous work we highlighted the need for supporting \textit{application-specific} intents, i.e., intents that are specific to the offerings of a website (e.g., ``Post a new comment on the news article") and that would trigger a series of low-level actions as a result. In our approach such experience required bot-specific annotations
\cite{chitto2020automatic}.
The automation of such workflows has also been explored in the context of Web accessibility. For example, Bigham et al. \cite{bigham2009trailblazer} introduced the trailblazer system, which focused on facilitating the process of creating web automation macros, by providing
step by step suggestions based on CoScript \cite{leshed2008coscripter}. It is also the focus of the research in robotic process automation \cite{mahala2020designing}.

\smallskip
\noindent\textbf{Conversational control operations.}
BVIP should be able to easily access and personalise the operational environment. 
This goes from simple operations to support the main browsing experience, such as searching and opening websites and managing the bookmarks, to personalising properties of the voice-based interactions. 
Recent works in this context have highlighted the importance of providing  BVIP with higher control over the experience. Abdolrahmani et al. \cite{abdolrahmani2018siri} investigated the experience by BVIP with voice-activated personal assistance and reported that users often feel responses being too verbose, frustrated at interacting at a lower pace than desired, or not able adapt interactions to the requirements of social situations.
It has been argued \cite{branham2019reading} that guidelines by major commercial voice-based assistants fail to capture preferences and experience of BVIP, used to faster and more efficient interactions with screen readers. This calls for further research into conversation design tailored to BVIP.

\subsection{Approaches and challenges}
There are many challenges in delivering the type of support required for conversational browsing. As discussed in our prior work \cite{chitto2020automatic}, this requires deriving two important types of knowledge:
\begin{itemize}
    \item \textbf{domain knowledge}: it refers to the knowledge about the type of functionality and content provided by the website, and that will inform the agent of what should be exposed to the users (e.g., intents, utterances and slots);
    
    \item \textbf{interaction knowledge}: it refers to the knowledge about how to operate and automate the browsing interactions on behalf of the user.
\end{itemize}

Websites are not equipped with the required conversational knowledge to enable voice-based interaction, which have motivated three general approaches.

The \textbf{annotation-based approach} provides conversational access to websites by enabling developers and content producers to provide appropriate annotations \cite{baez2019conversational}. 
Early approaches can be traced back to enabling access to web pages through telephone call services via VoiceXML \cite{oshry2007voice}. 
Another general approach to voice-based accessible information is to rely on accessibility technical specifications, such as Accessible Rich Internet Applications (WAI-ARIA) \cite{diggs2016accessible}, but these specifications are meant for screen reading. 
Baez et al. \cite{baez2019conversational} instead propose to equip websites with bot-specific annotations. 
The challenge in this regard is the adoption by annotations by developers and content producers. 
A recent report\footnote{https://webaim.org/projects/million/} analysing 1 million websites reported that a staggering 98.1\% of the websites analysed had detectable accessibility errors, illustrating the extent of the adoption of accessibility tags and proper design choices on the Web.

The \textbf{crowd-based approach} utilizes collaborative metadata augmentation approaches \cite{bigham2007accessmonkey,takagi2008social}, relying instead on the crowd to ``fix" accessibility problems or provide annotations for voice-based access. 
The Social Accessibility project \cite{sato2010exploratory} is one of these initiatives whose database supports various non visual browsers.
Still, collaborative approaches require a significant community to be viable, and even so the numbers of services and the rate at which they are created make it virtually impossible to cover all websites.

\textbf{Automatic approaches} have been used to support non-visual browsing and are based on heuristics and algorithms. 
The approaches in this space have focused on automatically fixing accessibility issues (e.g., page segmentation \cite{cormier2016purely}), deriving browsing context \cite{mahmud2007csurf} or predicting next user actions based on current context \cite{puzis2013predictive}. 
These approaches, however, have not focused on enabling conversational access to websites. 

\smallskip

All of the above tell us of the diverse approaches that can support the cognitive augmentation of websites to enable voice-based conversational browsing. 
In this work we explore automatic approaches, which have not been studied in the context of conversational access to websites. 

\section{A heuristic-based approach}

In this work we focus on heuristics that enable voice-based \textit{navigation} and \textit{access to content} of websites that are \textit{information intensive}. We focus on these set of features (Table \ref{tab:intents}, ``Browsing category") as they i) match the level of support expected but poorly served by screen readers, and  ii) are highly impacted by accessibility errors in websites.   

\subsection{Requirements}
From the conceptual framework, it becomes clear that enabling BVIP to browse websites conversationally requires us to:

\begin{itemize}
\item Determine the main (and contextual) offerings of the website
\item Identify the current navigation context
\item Enable navigation through meaningful segments of the website
\item Allow for scanning and search for information in the website
\end{itemize}

Determining the offerings of the website can be done by leveraging the components used in graphical user interfaces to guide users through their offerings: menus. Menus have specific semantic tags in HTML (\texttt{<nav>}) and roles as part of the technical specifications for web accessibility (WAI-ARIA) that allow screen readers to identify the main navigation links in a website. They also rely on distinctive visual and structural properties (e.g., styles and position) to make them easily identifiable by sighted users.  For example, they tend to be more prominent, towards the top and repeat across all pages in the website. Instead, more localised options are typically embedded in the content (e.g., links) or located within the same section of the page. 
Advanced models have relied in such visual properties to derive the role of rendered components in websites \cite{akpinar2017discovering}.

Identifying and keeping track of navigation context is supported in different ways by visual web browsing. In a website, this can be provided by design e.g., by implementing navigation breadcrumbs that explicitly render the navigation path that took users to their current context. It is also supported by Web browsers as part of the navigation history, allowing users to go back an forth in their navigation path but without illustrating it explicitly. In the context of a dialog, we can leverage this browsing history (available as a Web API) along with the conversation history to resolve the current browsing context based on navigation path (e.g., current page) and previous choices (e.g., name of links selected).  

Enabling navigation requires supporting browsing activities across different pages in the website, and therefore identifying relevant links and their target components. In visual browsing, the identification of the target components are typically done \textit{visually} by sighted users by relying on the layout of websites and their own goals. That is, when opening a news article, sighted users can focus their attention on the content of the article, ignoring other components such as headers, menus and ads. Given the proper accessibility tags, screen readers can allow users to (manually) identify their targets by skipping regions of the website. To provide a proper experience, it is fundamental to have meaningful segmentation of the website according to visual properties, and identifying target segments during navigation based on navigation context (e.g., as in done in \cite{mahmud2007csurf}). Segmentation techniques have been widely studied in accessibility research and could be leveraged for this purpose.

Searching within a website is not a particularly challenging feature. The challenge lies again in segmenting the resulting elements and contextualising the guidance based on the type of visual element (e.g., paragraph $\rightarrow$ reading; links $\rightarrow$ navigation). However, in accessibility this is associated to the \textit{non visual scanning} task, i.e., efficiently finding relevant information among many relevant ones, which has motivated several techniques, including the use of multiple concurrent audio channels \cite{guerreiro2016scanning}, that should be considered as potential techniques.

\subsection{Prototype implementation}
We implemented a prototype as an exercise into understanding and informing i) the type of support required in voice-based interactions, as well as ii) their technical requirements. The main focus was to establish a pipeline that can take voice commands, and fulfill them based on an (evolving) set of heuristics. In doing so, we faced the following architectural constraints: 

\begin{itemize}
\item The agent needs to serve multiple websites, as with a regular web browser
\item The agent needs to support conversational browsing intents, and the experience need to be optimised for ``reading" content
\item Processing times should be minimised for a meaningful user experience
\item The agent needs to support dynamic web pages
\end{itemize}

\begin{figure}[t!]
\centering
  \includegraphics[width=\columnwidth]{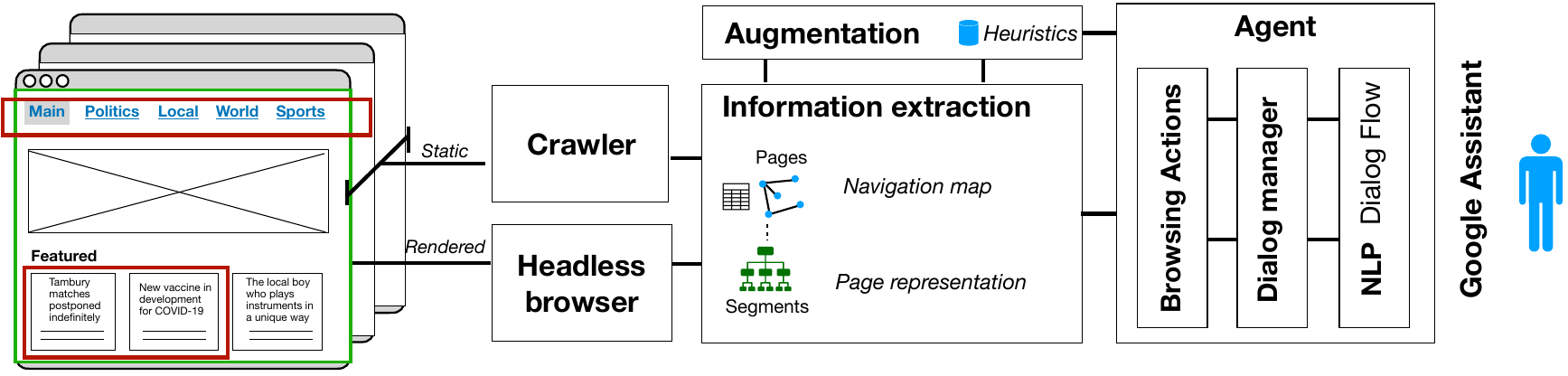}
  \caption{Pipeline for augmenting information-intensive websites with conversational capabilities by leveraging heuristics. }
  \label{fig:pipeline}
\end{figure}

The resulting pipeline is illustrated in Figure \ref{fig:pipeline}. In a nutshell, the pipeline takes a website and its static and dynamic content to create internal representations that can be leveraged by the heuristics to serve the predefined browsing intents. In the following we detail on this pipeline. 

\smallskip
\noindent\textbf{Crawling and data scraping.}
The first component in the pipeline is in charge of obtaining the static and dynamic contents and structure of the website for further processing and analysis. 
The process starts with the input URL to fetch the static HTML of each page in the website.  It performs a \textit{breadth-first search} of the website's tree structure,
identifying in each page all the hyperlinks to visit. This process is performed the first time the website is accessed, and it is cached (with an expiration date) for later use. The crawling runs in the background, stopping at a configurable depth $d$ in the tree or when a number of web pages $p$ have been processed. This process is implemented with \textit{Scrapy}\footnote{https://scrapy.org/}, a Python framework used for large scale web crawling.

While accessing the static version of a website ensures higher performance by reducing rendering times and allowing faster website-level analyses, it does not necessarily represent the actual content presented to the user, since part of it can be dynamically generated. For this reason, we complement the ``quick glance" provided by the crawling process by accessing the rendered version of the website on demand -- meaning the actual pages the user navigates to. The implementation relies on \textit{Selenium}, a powerful tool for automated testing in web browsers, running \textit{Mozilla Firefox} in headless mode to access the rendered version of the websites, with extensions such as AdBlocker and i-dont-care-about-cookies.eu to speed up rendering and loading times.

\smallskip
\noindent\textbf{Information extraction and augmentation.}
This component takes in input the \textit{website-level} information and the more detailed and accurate \textit{page-level} information from rendered pages to build internal  representations of the website.
The website-level information is leveraged to build the navigation graph of the website, and calculate basic metrics on the structure (e.g., popularity: the number of times a link is referenced) that can later inform the heuristics. Basic metadata is extracted but the static HTML is not further processed at this stage.
Then, when the user navigates to specific page (a node in the navigation graph) the rendered version of the website is requested, and the actual content and structural properties of the website are analysed. The page is represented as a tree-structure, much like the DOM but where each node is a meaningful segment of the website, as derived by the segmentation heuristics (described later). The contents of the nodes are extracted and cleaned to make them reading friendly (e.g., inline links replaced by placeholders and offered separately).
The implementation of this component relied on the \textit{BeautifulSoup}\footnote{https://www.crummy.com/software/BeautifulSoup/bs4/doc/} Python package to analyse the HTML code and scrape data.



\smallskip
\noindent\textbf{Computing heuristics.}
The current prototype implements simple heuristics that serve as placeholders that will allow more comprehensive tests of the entire pipeline. An example of such heuristics, for the identification of the main offerings of the website, is based on the observation that links in the main menu tend to be at the top of the page and present across the entire website. We therefore leveraged on the navigation map and the calculated popularity metric for each node (i.e., how many times the link is referenced), weighted by the position attribute of the link element in the rendered website (e.g., thus discerning links in footer and headers) to rank the links. We do not currently perform segmentation, and the segmentation placeholder just leverages existing region landmarks.  

Other features of Table \ref{tab:intents} currently rely on existing cognitive services. For example, for providing \textit{Summary}, we rely on Aylien Text Analysis API\footnote{\url{https://fortiguard.com/webfilter}}, which along with  Fortiguard Web Filter\footnote{https://fortiguard.com/webfilter} augments the information about the website with extra metadata (e.g., language and topic of the website). The search feature are provided by Google Search. 


\subsubsection{Conversational Agent.}
The browsing experience is ultimately delivered through \textit{Google Assistant}, which was chosen as the voice-based service. This service provides a conversational medium and performs the speech-to-text and text-to-speech transformations to and from the natural language processing unit. 
We relied on \textit{Dialogflow}\footnote{https://dialogflow.com/} as natural language platform, where the intents for serving the conversational browsing needs were defined. These include the Browsing, and a few of Operations and Metadata \& Content from Table \ref{tab:intents}. 
The webhooks to handle the fulfillments ultimately pointed to our Python server. The source code of our prototype is available at \url{https://github.com/Shakk17/WebsiteReader}.

\subsection{Preliminary Evaluation}
A preliminary evaluation of the system was performed so as to assess the technical performance of the tool, and gain insights on the structure of websites and the challenges they present to our heuristic-based approach. 

A total of 30 websites were selected from Alexa's top ranking\footnote{https://www.alexa.com/topsites/category/Top}, taking 5 websites from each of the six categories typically associated with information-intensive websites: Newspapers,
Sports, Reference, Health, Society and Science. We tested the accessibility compliance of these top websites with the WAVE accessibility tool\footnote{https://wave.webaim.org}. This revealed that only 4 out of the 30 websites were free of accessibility errors, which further illustrates the challenges to our approach and to assistive technology in general.  

In this exploratory run, we evaluated the performance of the simple heuristic for inferring the offerings of the website. To do this, we first manually analysed each website to identify the links from the menus (the offerings). These actual links were then compared against the output of the heuristic, which was we set to return a maximum of 30 links (threshold), to compute precision and recall. 

The results showed that the heuristic is effective in identifying relevant links (recall=0.79) but less precise in determining the number of links to recommend (precision=0.42). However, this is mainly due to the static threshold and the highly wide range of menu size and complexity in websites (from 4 to 40 links). Indeed, the precision was much higher when the number of recommended links approached the number of actual links in the menu.


Our observations running these tests tell us that the solution goes beyond more intelligent heuristics and cut-off values for links. 
The analysis revealed the complexity of menus in websites -- some of then with dozens of hierarchical links -- which motivates an exploration into new approaches to presenting and discovering available offerings conversationally. The exploration of conversational patterns for menu-access as well as heuristics for identifying global and local intents (links) emerge are interesting areas for future research.

\section{Discussion and Future Work}

In this paper we have explored the opportunities of cognitive augmentation and automation to support BVIP in browsing information-intensive websites. The approach is based on the notion of enabling dialog-based interaction with websites, mediated by a voice-based conversational agent. 

These opportunities were materialised in a conceptual framework that summarised based on literature review, our prior work and  prototyping exercises, the categories of support to be addressed to enable conversational browsing by BVIP. These include the ability to interact with the contents of the website, support more traditional browsing tasks, automating user workflows and managing the entire operating environment of the browsing experience. The infrastructure of the Web today is not equipped serve these needs, but we have shown that cognitive computing can enable and augment the existing foundation -- much as with cognitive process augmentation \cite{barukh2020} -- to help address these needs. Existing research and techniques in accessibility can greatly kick-start these efforts. 

It is however clear that automation alone cannot fulfill this vision. Delivering a proper conversational experience, under the limitations and constraints posed by the problem, would require addressing technical issues as well as implementing dialog patterns that can reduce their impact and provide guidance. Equipping website with conversational knowledge, reflecting the intended conversational experience, appears to be key in this regard. Understanding the correct trade-off between what should be explicitly annotated and can be automatically derived are among the challenges to be addressed. 

As part of our ongoing work we are planning to integrate a pool of existing algorithms and heuristics developed in the accessibility community and setup benchmarks to understand their suitability and performance. We are also planning user studies to understand the impact of different dialog patterns, associated with different levels of explicit and implicit conversational knowledge. 
The long-term vision is to integrate conversational capabilities into systems of any kind, a problem that we have seen emerging and  gaining traction in recent years
\cite{baez2020chatbot}.

\bigskip
\noindent \textbf{Acknowledgements}. This work was carried out with the precious contribution from our friend and mentor, Florian Daniel, who passed away shortly before completing this project. We remember him here \cite{baez2020remembering}.

%
%
\bibliographystyle{splncs04}
\bibliography{references}

\end{document}